\documentstyle[12pt,thmsa]{article}
\input{tcilatex}
\begin{document}

\begin{center}
{\LARGE Thermal properties of ferrimagnetic systems}

{\Large \ }\vspace{1cm}

Aiman Al-Omari \cite{mail} and A. H. Nayyar\cite{mail1}\\[0pt]

{\it Department of Physics}, \\[0pt]
{\it Quaid-i-Azam University, }\\[0pt]
{\it Islamabad, Pakistan 45320}

\vspace{0.7cm}(January 22, 2000)\vspace{1.2cm}
\end{center}

The heat capacity of some ferrimagnets has additional structures like a
shoulder in the Schottky-like peak, or emergence of a second peak when an
external magnetic field is applied. It is shown here that as long as spin
wave-spin wave interactions are ignored in a ferrimagnet, the ferromagnetic
and antiferromagnetic elementary excitation spectra give rise to two
independent heat capacity peaks, one enveloped by the other, which add up to
give the peak for the total system. Taking this into account helps
understand the additional structures in the peaks. Moreover, the
classification of ferrimagnets into predominantly antiferromagnetic,
ferromagnetic, or a mixture of the two is shown to be validated by studying
them under additional influences like dimerization and frustration. Because
these two are shown to influence the ferromagnetic and antiferromagnetic
dispersion relations - and hence the quantities like heat capacity and
magnetic susceptibility - by different amounts, the characterisation of
ferrimagnetic systems ($1,\frac{1}{2}$), ($\frac{3}{2},1$) and ($\frac{3}{2},%
\frac{1}{2}$) is brought out more clearly. Both these influences enhance
antiferromagnetic character.\newline

\noindent PACS numbers: 75.10.Jm, 75.50.Ee\vspace{0.5cm}\newpage

\section{\strut Introduction:}

Thermal properties of ferrimagnetic chains were theoretically investigated
recently employing various methods like the modified spin-wave theory (MSWT)%
\cite{yamamoto,yamamoto2}, density matrix renormalization group (DMRG)\cite
{yamamoto2}, quantum Monte Carlo method (QMC)\cite{yamamoto,yamamoto2}, and
Schwinger boson mead field theory (SB)\cite{wu}. The specific heat $C_{v}$
and magnetic susceptibility $\chi $ were shown to depend upon temperature as
T$^{1/2}$ and T$^{-2}$ respectively at low temperatures, and $\chi T$ and $%
C_{v}$ were shown to have, respectively, a rounded minimum\cite
{yamamoto2,wu,pati} and a Schottky-like peak\cite{yamamoto2,pati} at
intermediate temperatures. The spin correlation length was shown to have a $%
T^{-1}$dependence at low temperature\cite{yamamoto,yamamoto3}. The modified
spin wave theory, modified either by including Takahashi constraint or by
including higher order corrections in the spin wave theory, was also shown
to give results in surprisingly good agreement with those from quantum Monte
Carlo method in the thermodynamic limit for this system\cite
{yamamoto,yamamoto2}.

Ferrimagnetic systems have been classified into three categories: one with a
predominantly Ferromagnetic (F) character, the second with a predominantly
antiferromagnetic (AF) character, and the third with a mixture of the two.
If the two spins constituting a ferrimagnet are $s_{1}$ and $s_{2}$ with $%
s_{1}>s_{2}$, then it is conjectured\cite{yamamoto4,yamamoto5} that the
systems in the first category are those with $s_{1}>2s_{2}$, those in the
second category have $s_{1}<2s_{2}$, and the systems in the third category
have $s_{1}=2s_{2}$. We call it Yamamoto classification{\bf .} Thus the
system $(\frac{3}{2},\frac{1}{2})$ is ferromagnetic in character, $(\frac{3}{%
2},1)$ is antiferromagnetic, and $(1,\frac{1}{2})$ has a mixture of the two
characters.

The $\chi T$ vs temperature curves of different ferrimagnetic systems look
alike - a rapidly decaying ferromagnetic part at low $T$, a rounded minimum
at intermediate $T$ and a linearly increasing antiferromagnetic part at high 
$T$ - except that some systems like the predominately ferromagnetic $(\frac{3%
}{2},\frac{1}{2})$ have a smaller rate of increase with temperature after
the minimum compared to others. The heat capacities of the three systems
also have qualitatively the same shape:\ a $T^{1/2}$ dependence at low $T$,
a Schottky-like peak and a decay at large temperatures. The MSW results on $(%
\frac{3}{2},\frac{1}{2})$ system, however, show a shoulder in the heat
capacity below the peak temperature, which has been explained as being a
result of the deficiency of the theory\cite{yamamoto5}. When an external
magnetic field is applied, a second peak appears at low temperatures\cite
{maisinger}. It is, therefore, quite likely that the heat capacity of
ferrimagnets has, under suitable conditions, inherent structures like
shoulders and double peaks. We would like in this paper to create such
conditions and understand the nature of the heat capacity peak of
ferrimagnets.

\smallskip In a recent work\cite{aiman1}, we used a zero-temperature linear
spin wave theory to study ferrimagnetic systems under the effects of
dimerization and frustration, represented by the parameters $\delta $ and $%
\alpha $. Dimerization as well as the frustration were shown to affect the
dispersion curves by either pushing them up or pulling them down in energy.
Since this would change the gap that determines the heat capacity peak,
these two effects are expected to influence the peak structure. It is also
possible that they may give rise to additional peaks in heat capacity.

There were two distinct values of the frustration parameter; the critical
value $\alpha _{c}=\frac{s_{1}}{2(s_{1}+s_{2})}$ at which the long range
order is destroyed, and point $\alpha ^{*}=\frac{s_{1}s_{2}}{%
2(s_{1}^{2}+s_{2}^{2})}$, that heralds transition from a commensurate
ferrimagnetic state to a spiral state. The heat capacity and susceptibility
are also expected to show the tell-tale signs of the transition induced by $%
\alpha $.

Linear spin wave theory is known to give a fair picture in the case of
ferromagnetic chains and gives only a qualitative picture for
antiferromagnetic chains \cite{yamamoto5}. It has also been shown that it
gives sufficiently good results for a ferrimagnetic system\cite
{brehmer,yamamoto57,ivanov,ivanov57}. In a frustrated system, it has already
been argued that the LSW theory yields satisfactory results at least in the
limit of small frustration\cite{ivanov}. The use of LSW for larger values of
frustration is indeed unreliable, but it is expected to give a qualitative
picture that we are seeking here. As noted above, there was no remarkable
improvement in the results at non-zero temperatures when linear spin wave
theory was modified either by introducing a Takahashi constraint\cite
{yamamoto5} or by including higher order corrections in the spin wave theory%
\cite{yamamoto,yamamoto2}. Nor were the results at zero temperature any more
improved by considering non-linear spin wave theory\cite{aiman2}. The LSW
theory is therefore expected to be valid in obtaining qualitative results
even in the presence of dimerization and frustration in the thermodynamic
limit.

In this paper we will study alternating spin systems formed with spin values
($s_{1},s_{2})$ using linear spin wave theory. We investigate the
temperature dependence of magnetization, specific heat, free energy, and
susceptibility. We would like to see how the predominantly ferromagnetic,
antiferromagnetic or mixed characters of the systems are brought out by
subjecting them to dimerization and frustration. In particular, we would
like to see the effect of the frustration-induced ferrimagnetic-to-spiral
state phase transition on heat capacity and magnetic susceptibility. It is
suggested that such peculiar effects will help identify the presence or
absence of spin-Peierls dimerization and frustration in real low-dimensional
ferrimagnetic systems.

\section{Linear spin wave theory:}

We consider a chain with spins $s_{1}$ and $s_{2}$ ( $s_{1}$ $>$ $s_{2}$)
sitting on alternating sites with the possibility of lattice distortion
leading to dimerization, and of competing antiferromagnetic nearest and next
nearest neighbor couplings, $J_{1}$ and $J_{2}$ respectively. A two
sublattice model of this system may be described by the Hamiltonian 
\begin{equation}
H=\sum_{i}J_{1}^{i}S_{i}^{A}\cdot S_{i+1}^{B}+J_{2}\sum_{i}\left[
S_{i}^{A}\cdot S_{i+2}^{A}+S_{i+1}^{B}\cdot S_{i+3}^{B}\right] ,
\label{1dham}
\end{equation}
with $S^{A}$ belonging to one sublattice containing spins $s_{1}$ and $S^{B}$
to the second sublattice containing spins $s_{2}$. The interaction $%
J_{1}^{i}=J_{1}(1+$ $(-1)^{i}\delta )$ describes the alternate weaker and
stronger couplings between two adjacent sites that may come about because of
a spin-Peierls dimerization of the lattice. $\delta $ is the diemrization
parameter. The degree of frustration is given by the ratio $\alpha =\frac{%
J_{2}}{J_{1}}.$

In the standard linear, non-interacting spin wave analysis, the above
Hamiltonian is written in terms of bosonic spin-deviation operators with the
help of Holstein-Primakoff transformations, linearized and then diagonalized
in terms of normal mode operators to

\begin{equation}
\tilde{H}=\varepsilon _{g}+\sum_{k}\left[ E_{1}(k)\alpha _{k}^{\dagger
}\alpha _{k}+E_{2}(k)\beta _{k}^{\dagger }\beta _{k}\right]
\end{equation}
with $\varepsilon _{g}$ is the ground state energy, $E_{1,2}$ are the
acoustic and optical mode energies and the sum runs over half the Brillouin
zone. The nomenclature used here is the same as in our previous report\cite
{aiman1}.

The free energy of the system is 
\begin{equation}
F=\varepsilon _{g}+\sum_{k}[n_{a}E_{1}(k)+n_{b}E_{2}(k)],  \label{feng}
\end{equation}
where $n_{a}$ and $n_{b}$ are the bose distribution functions for the two
modes. It was reported earlier that the free energy decreases with
increasing temperature\cite{wu} in a ferrimagnetic chain.

The static susceptibility $\chi $ is 
\begin{eqnarray}
\chi &=&\frac{1}{3T}\left( <M^{2}>-<M>^{2}\right)  \label{sus1} \\
&=&\frac{1}{3T}\sum_{k}[n_{a}(n_{a}+1)+n_{b}(n_{b}+1)]
\end{eqnarray}
where $M=\sum\limits_{i}(S_{a}^{z}+S_{b}^{z})$. The product $\chi T$ shows a
minimum at intermediate temperatures, indicating that both the ferromagnetic
and antiferromagnetic modes coexist in ferrimagnets.

The sublattice magnetization is calculated by taking the thermal average of
the occupation number as 
\begin{eqnarray}
M_{1} &=&s_{1}-\sum_{k}\widetilde{n}_{k}^{a} \\
M_{2} &=&\sum_{k}\widetilde{n}_{k}^{b}-s_{2}
\end{eqnarray}
and 
\begin{equation}
M_{tot}=M_{1}+M_{2}=s_{1}-s_{2}-\sum_{k}\left( n_{a}-n_{b}\right)
\label{totmg}
\end{equation}
where $\widetilde{n}_{k}^{a}=v_{k}^{2}(n^{a}+n^{b}+1)+n^{a}$ , $\widetilde{n}%
_{k}^{b}=v_{k}^{2}(n^{a}+n^{b}+1)+n^{b}$ and $v_{k}^{2}$ is the coefficient
in the Bogoliubov transformation; see Ref.\cite{aiman1}.

Specific heat $C_{v}$ has a Schottky-like peak. In a text-book two level
system, the peak is a result of the gap, and an increase in this gap makes
the peak shift to higher values of temperature, and a decrease makes it
shift to lower values.

A ferrimagnet has two low lying excitations, one ferromagnetic in character
(the acoustic dispersion) and the other antiferromagnetic (the optic
dispersion curve). The Schottky like peak in the heat capacity is understood
to be due to the two dispersion curves. In principle, because of the two
excitation dispersion curves, one would expect this to behave like a three
level system. But the heat capacity of a ferrimagnet does not follow the
pattern of a 3 level system. Instead, from the non-interacting spin wave
theory results of ferrimagnets, the contributions of the ferromagnetic and
antiferromagnetic dispersions to the heat capacity of a ferrimagnet seem to
be additive. The heat capacities calculated for the modes separately has
each a Schottky-like peak structure, and the two add up to give the heat
capacity of the ferrimagnet. This is also expected from the fact that in the
absence of spin wave-spin wave interactions, the free energy is a sum of the
ferromagnetic (F) and antiferromagnetic (AF) normal mode energies, as in Eq.(%
\ref{feng}). The F and AF heat capacity peaks, which have the usual $T^{1/2}$
and $T^{2}$ dependence, respectively, at low temperatures, are mostly
overlapping due to which the resultant heat capacity has a single peak. The
additive nature of the two contributions is also reflected in the dependence
of susceptibility on temperature. With increasing temperature, the
ferromagnetic contribution decays earlier, while the antiferromagnetic one
increases and persists up to higher temperatures \cite
{yamamoto,yamamoto2,wu,pati}

The behavior of these physical quantities with temperature in the presence
of dimerization $\delta $ and frustration $\alpha $ parameters for different
system ($s_{1},s_{2});(1,\frac{1}{2}),$ $(\frac{3}{2},\frac{1}{2})$ and ($%
\frac{3}{2},1),$ shall be discussed below. The effect of dimerization alone
is discussed in section III, and that of frustrating alone in section IV.

\section{Dimerized chains}

The free energy is known to decrease with temperature. We find that it
decreases with dimerization also for all ($s_{1},s_{2}$) systems, as shown
for the $(1,\frac{1}{2})$ system in Fig.(1). The LSW results show that the
free energy scales as $T^{1/4}$, and{\bf \ }$\delta $ as $\delta ^{2}$.{\bf %
\ }

Total magnetization decreases with increasing $T$ and $\delta ,$ with a
slower decrease for higher dimerization. This decrease is illustrated in
Fig.(2), and scales as $T$ , and{\bf \ }$\delta $ as $\frac{\delta ^{1.5}}{%
|\ln \delta |}$.{\bf \ }

Fig.(3) shows the effect of dimerization on the heat capacity of the (1,$%
\frac{1}{2})$ ferrimagnetic chain which is supposed to have a mixed F and AF
characters. Fig.(3a) gives the heat capacity without dimerization, and
Fig.(3b) with $\delta =0.9.$ The figures show the heat capacities when the
contributions of the ferromagnetic and antiferromagnetic excitation branches
are taken separately and also when taken together. As expected, the total
heat capacity is the sum of the heat capacities from the individual
excitation branches in each case. It comes out that while both the F and AF
peaks shift to lower temperatures with dimerization, the effect is different
on the two. The F peak shifts by a larger amount than the AF peak, and the
magnitude of the AF peak increases with $\delta ,$ while that of the F peak
decreases. This is true for the other two ferrimagnetic systems ($\frac{3}{2}%
,1$) and ($\frac{3}{2},\frac{1}{2}$) also, and appears to be a universal
feature of a ferrimagnet, be it predominantly F, AF or a mixture the two
characters. The increase and decrease in the AF and F peak magnitudes
clearly shows that dimerization increases the antiferromagnetic character
and decreases the ferromagnetic one. The net effect of this different effect
is the emergence of the second peak in the total heat capacity, and a
decrease in the magnitude of the main peak for the (1,$\frac{1}{2})$ system,
as shown in the inset. The predominantly F system ($\frac{3}{2},\frac{1}{2}$%
) has this effect so accentuated that the shoulder, which in the LSW and MSW
analyses is present even at $\delta =0$, turns into a double peak, as shown
in Fig.(4a). The predominantly AF system ($\frac{3}{2},1$), on the other
hand, shows no second peak or even a shoulder, as in Fig.(4b). It is
interesting to note that the difference arises despite the fact that the F
and AF dispersion curves are lowered by an equal amount by dimerization. It
appears to be a result of a more involved interplay of the $\delta $%
-dependence of the excitation energies and the bosonic numbers in the free
energy.

The shift of the peaks to lower temperatures can be understood in terms of
the energies of excitation. The heat capacity peaks are indeed Schottky-like
peaks in that they are a result of the gaps in excitation spectra. The
Schottky peak in the heat capacity of a simple text-book two-level system
with a gap $\Delta $ between the two levels has a position that shifts with
the magnitude of $\Delta $. As $\Delta $ decreases, it shifts to lower
temperatures, and vice versa. For a ferrimagnetic chain, contribution to the
peak comes from all the modes which have well-defined dispersions in
k-space, each mode having a different gap. With dimerization, the dispersion
curves are lowered in energy, giving rise to a shift of the peak to lower
temperatures.

That the shifting is a result of the changes in the excitation spectra is
also supported when the effect of dimerization on heat capacity is
calculated by using a different parametrization of dimerization. We had
earlier introduced $\frac{J}{1\mp \delta }$ as another possibility\cite
{aiman1,aiman}. The reasons are given in that reference. This
parametrization has an opposite effect on the dispersion curves. Instead of
decreasing in energy with $\delta $ as in the case of the usual $J(1\pm
\delta )$, the dispersion curves are now pushed up in energy with increasing 
$\delta $, with a nonlinear dependence on $\delta .$ Fig.(5) shows the heat
capacity of the (1,$\frac{1}{2})$ chain with this parametrization. The
Schottky peaks now shift to higher temperatures with increasing $\delta $.
Also, the individual F and AF peaks respond differently to $\delta $. Just
as the rate of increase of the AF dispersion curve with $\delta $ is much
larger at higher $\delta $ values than that of the F dispersions, the AF
Schottky peak shifts by a much larger amount than the F peak. Consequently,
there comes a time when the AF peak is pushed so far to the high temperature
side that a double peak structure develops. This appears more clearly in the
case of the ($\frac{3}{2},\frac{1}{2})$ ferrimagnetic chain in which
ferromagnetic character dominates. This is shown in Fig.(6). The AF peak is
so strongly pushed to higher temperatures that the F and AF peaks almost
separate from each other. Correspondingly, in the dispersion curves of this
system, the optic mode is pushed up by a far larger magnitude than the
acoustic one. This establishes then that the heat capacity peaks reflect the
effects an external or internal influence has on the dispersion curves.

The change of character upon dimerization shows up in the susceptibility of
ferrimagnets also. Fig.(7) shows $\chi T$ vs $T$, the low temperature part
of which is dominated by the F contribution and the high temperature part by
the AF contribution. The minimum at intermediate temperatures is a result of
a comparable contributions of the two. With dimerization using{\bf \ }$%
J(1\pm \delta )${\bf ,} the increase in the slope at high temperatures
testifies to the increasing AF character of a ferrimagnet. This increase is
the largest for the predominantly AF ($\frac{3}{2},1)$ system; Fig.(7c),
smaller for the ($1,\frac{1}{2})$ system; Fig.(7a){\bf \ }and the smallest
for the predominantly ferromagnetic ($\frac{3}{2},\frac{1}{2}$) system;
Fig.(7b), showing again the validity of the Yamamoto classification.

\section{Frustrated ferrimagnetic chains}

The effect of frustration in ferrimagnetic systems has been investigated\cite
{aiman1,ivanov} at zero temperature. Two key values of the frustration
parameter were identified; namely, $\alpha _{c}$ which marks complete
destruction of the long range order in the system and $\alpha ^{*}$ which
marks the onset of phase transition to a spiral spin structure. At any
finite temperature, we find that within the linear spin wave theory the two
are temperature independent. We will study the thermal effect of a
frustrated chain in the absence of dimerization.

Free energy decreases with frustration. This decrease is faster for higher $%
\alpha $. It was reported earlier\cite{aiman1,ivanov} that at $T=0$ the
ground states energy increases with frustration and then starts to decrease
for values of frustration close to $\alpha _{c}$. This behavior is
illustrated in Fig.(8) for the spin system $(1,\frac{1}{2})$.

The effect of frustration on heat capacity will again be discussed for the F
and AF contributions separately. Frustration causes the AF $C_{v}$\ peak to
slightly increase in size and shift to lower temperatures. But its effect on
the $C_{v}$\ peak due to the F excitation modes is more dramatic. Here the
transition to the spiral phase at $\alpha ^{*}$\ has its impact. Before $%
\alpha ^{*}$\ the peak decreases in size and shifts to lower temperatures.
Beyond this value, when the system goes into the spiral phase, a further
increase in frustration causes the peak to increase in size and shift to
higher temperatures. All of this is a result of the mode-softening changes
that the F dispersion curves go through, as seen in Fig.(9). The total $C_{v}
$\ being a sum of the AF and F contributions in the non-interacting
spin-wave theory changes with $\alpha $\ as shown in Fig.(10).

In the case of the ($1,\frac{1}{2}$) system, the $\alpha =0$ chain having a
mixed F and AF characters, the $C_{v}$ curve has an AF-like $T^{2}$\
dependence at low temperatures. With added frustration, the F peak separates
out at low temperatures making the total $C_{v}$ curve have an F-like $%
T^{1/2}$\ dependence at low $T$. The F peak diminishes in size with
increasing $\alpha $, almost vanishing at $\alpha ^{*}$, beyond\ which it
regrows and eventually merges with the AF peak. At this point the $C_{v}$
curve again has an AF-like $T^{2}$ dependence at low temperatures.\ This is
shown in Fig.(10a)

In\ the case of the predominantly F system ($\frac{3}{2},\frac{1}{2}$), the
heat capacity of which has a low temperature shoulder due to the F peak even
in the absence of frustration, the changes affecting the F peak are very
prominent, as in Fig.(10b). As $\alpha $ increases, the F peak separates out
and diminish in size before $\alpha ^{*}$, after which\ it rises and
eventually merges with the AF peak. The net $C_{v}$ also experiences a
low-temperature cross-over from a ferromagnetic $T^{1/2}$ dependence at zero
frustration to the antiferromagnetic $T^{2}$\ dependence at $\alpha $ beyond 
$\alpha ^{*}$ in the spiral state.

In the case of the predominantly AF system ($\frac{3}{2},1$), the F peak in
itself experiences the same changes as in the ($1,\frac{1}{2}$) system, but
it remains submerged in the AF peak, and the net $C_{v}$ never shows a
second peak, as in Fig.(10c). It, however, also experiences the cross-over
from $T^{2}$ to $T^{1/2}$ behavior at low temperatures.

The effect of frustration on $\chi $ is the same as that of dimerization.
The minimum shifts to lower temperatures and the slope of the high
temperature tail increases as shown in Fig.(11).

In short, the effect of dimerization and frustration on the thermal behavior
provides a nice evidence of the validity of the Yamamoto classification of
ferrimagnetic chains.

We thus conclude that the heat hapacity of a ferrimagnet can have a more
complicated structure than a simple Schottky-like peak. It can have a low
temperature shoulder or a double peak structure. The peak can shift to lower
or higher temperatures and can also change in size depending upon the
influences the system is undergoing. All of this can be understood in terms
of the component ferromagnetic and antiferromagnetic peaks that retain their
individuality as long as the spin wave-soin wave interactions are ignored.
The Yamamoto classification of the ferrimagnetic systems into predominantly
ferromagnetic ($s_{1}>2s_{2})$, antiferromagnetic ($s_{1}<2s_{2})$ and
ferrimagnetic ($s_{1}=2s_{2})$ appears to have a basis in this
characteristic.

A. A. would like to thank Wu Congjun for discussion and help.

\newpage

{\bf Figure captions}\newline

Figure 1: Dependence of the free energy on $T$ for several values of the
dimerization parameter $\delta $ in one-dimensional spin system (1,$\frac{1}{%
2}$)$.$ A similar behavior was found for the other two systems $(\frac{3}{2},%
\frac{1}{2}$) and ($\frac{3}{2},1)$.

Figure 2: The total magnetization of the alternating spin chain (1,$\frac{1}{%
2}$) vs temperature for different values of the{\bf \ }dimerization
parameter $\delta .$ A systematic \hspace{0in}behavior was observed for the
other two systems $(\frac{3}{2},\frac{1}{2}$) and ($\frac{3}{2},1)$.

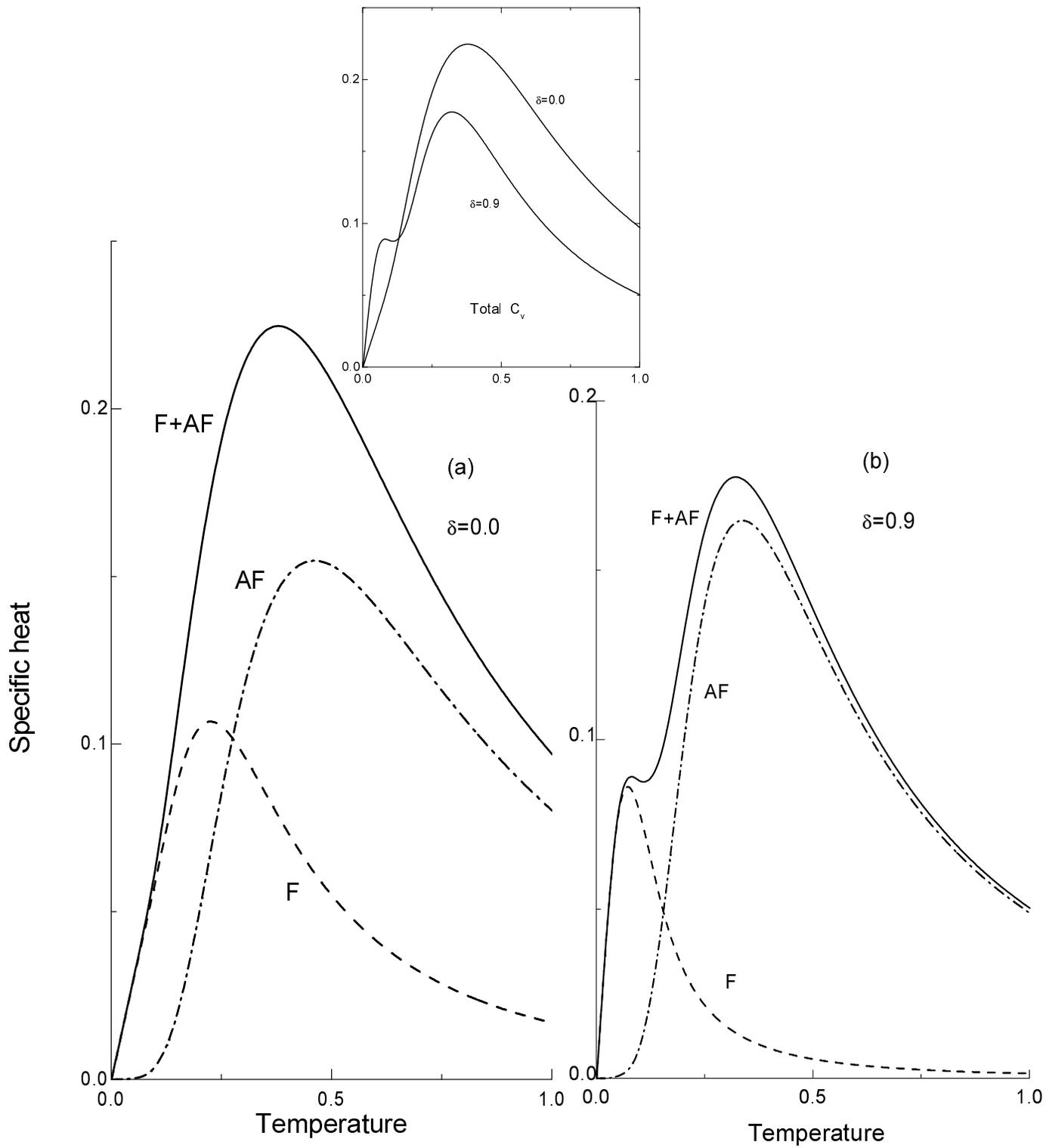
Figure 3: Heat capacity for the chain (1,$\frac{1}{2}$) as a function of $T$
. The heat capacities calculated with only the ferromagnetic (acoustic)
branch of the elementary excitation spectrum and with only the
antiferromagnetic (optic) branch are shown separately. Their sum is also
shown which is exactly the heat capacity when the two excitation branches
are taken into account together. Figure (a) is for $\delta =0.0$, and (b)
for $\delta =0.9$. The inset for total $C_{v}$ for the two values of $\delta 
$ highlights how the Schottky-like peak shifts in temperature, reduces in
size, and acquires a shoulder. The curves are shown for the system (1,$\frac{%
1}{2}$).

Figure 4: (a) Heat capacity of the ferrimagnetic chain $(\frac{3}{2},\frac{1%
}{2}$). This being a predominantly F system, the Schottky-like peak has a
shoulder even in the absence of dimerization. The effect of dimerization is
strong enough to separate out the F and AF peaks to result in a two-peak
structure. Fig.(b) is for the predominantly AF system ($\frac{3}{2},1)$
which fails to show a shoulder or a double peak structure even at large
dimerization.

Figure 5: The dependence of heat capacity on $T$ when the dimrization is
parametrized by the interaction $\frac{J_{1}}{1\mp \delta }$ for the
alternating spin chain (1,$\frac{1}{2}$). The shift in the Schottky-like
peak is to higher $T$, in contrast to the shift to lower $T$ in Fig.(3). A
large dimerization pushes the AF peak much farther than the F peak,
resulting in a two-peak structure.

Figure 6: The heat capacity for the alternating spin chain ($\frac{3}{2}$,$%
\frac{1}{2}$) with the parametrization as in Fig. (5). The F and AF peaks
are now very dramatically separated.

Figure 7: The $\chi T$ vs $T$ in ferrimagnetic chains: (a) for the system (1,%
$\frac{1}{2}$), (b) for $(\frac{3}{2},\frac{1}{2}$) and (c) for ($\frac{3}{2}%
,1),$ for various values of the dimerization parameter $\delta $. That
dimerization enhances the AF character of ferrimagnets is clearly observable
here.

Figure 8: The free energy for the alternating spin chain (1,$\frac{1}{2}$)
as a function of temperature for several values of the frustration parameter 
$\alpha $. The same behavior was found for $(\frac{3}{2},\frac{1}{2}$) and ($%
\frac{3}{2},1)$.

Figure 9: The dependence of the excitation energies on frustration parameter 
$\alpha $ for the spin system (1,$\frac{1}{2}$) shows the mode softening
that heralds the transition to spiral state.

Figure 10: Specific heat vs $T$ : (a) for the spin system (1,$\frac{1}{2}$),
(b) for ($\frac{3}{2},\frac{1}{2})$ and c) for $(\frac{3}{2},1$)$,$ for
various values of the frustration parameter $\alpha $.

Figure 11: The behavior of $\chi T$ vs $T$ of the ferrimagnetic system (1,$%
\frac{1}{2}$) on a chain for different values of $\alpha $.

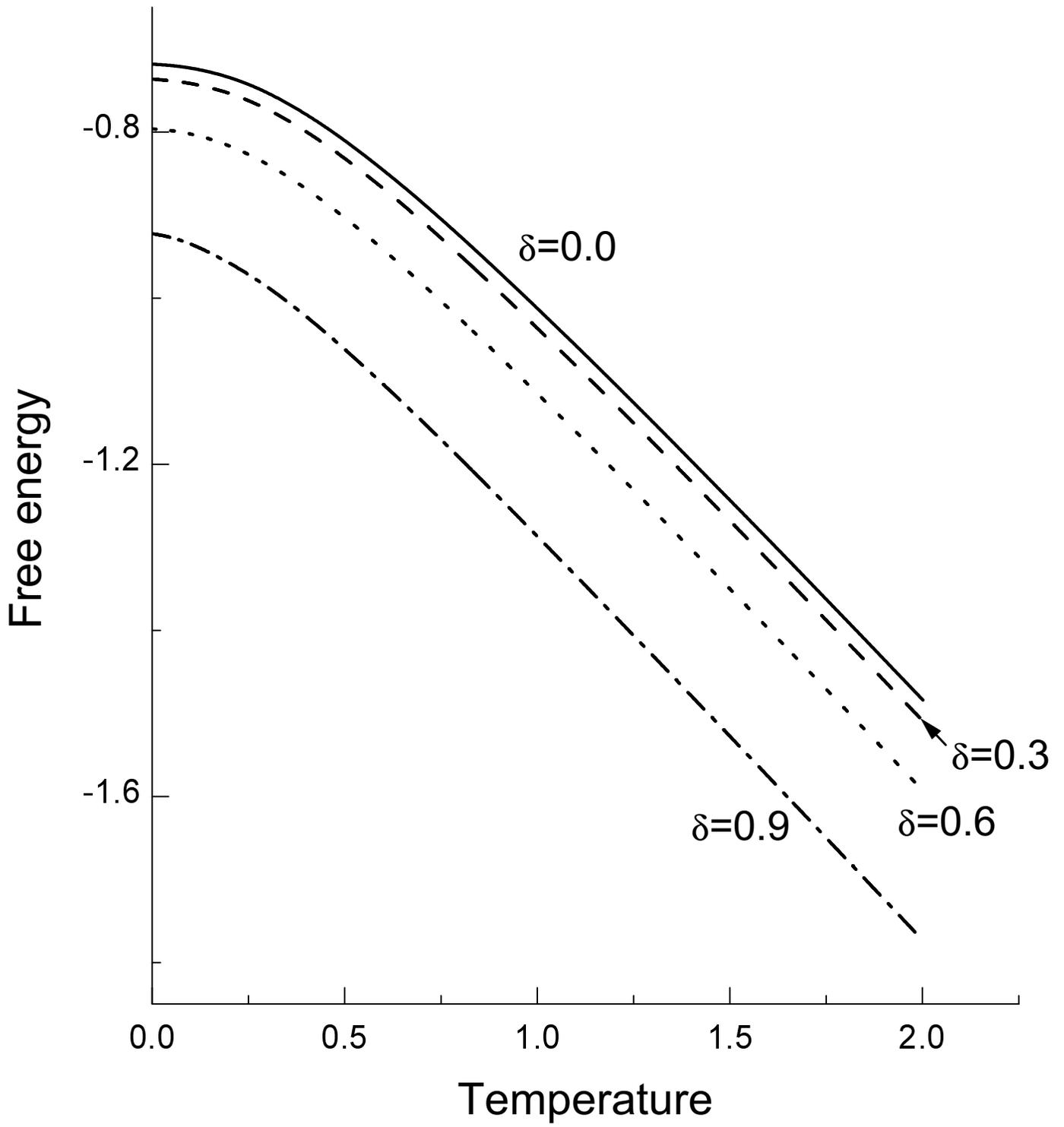
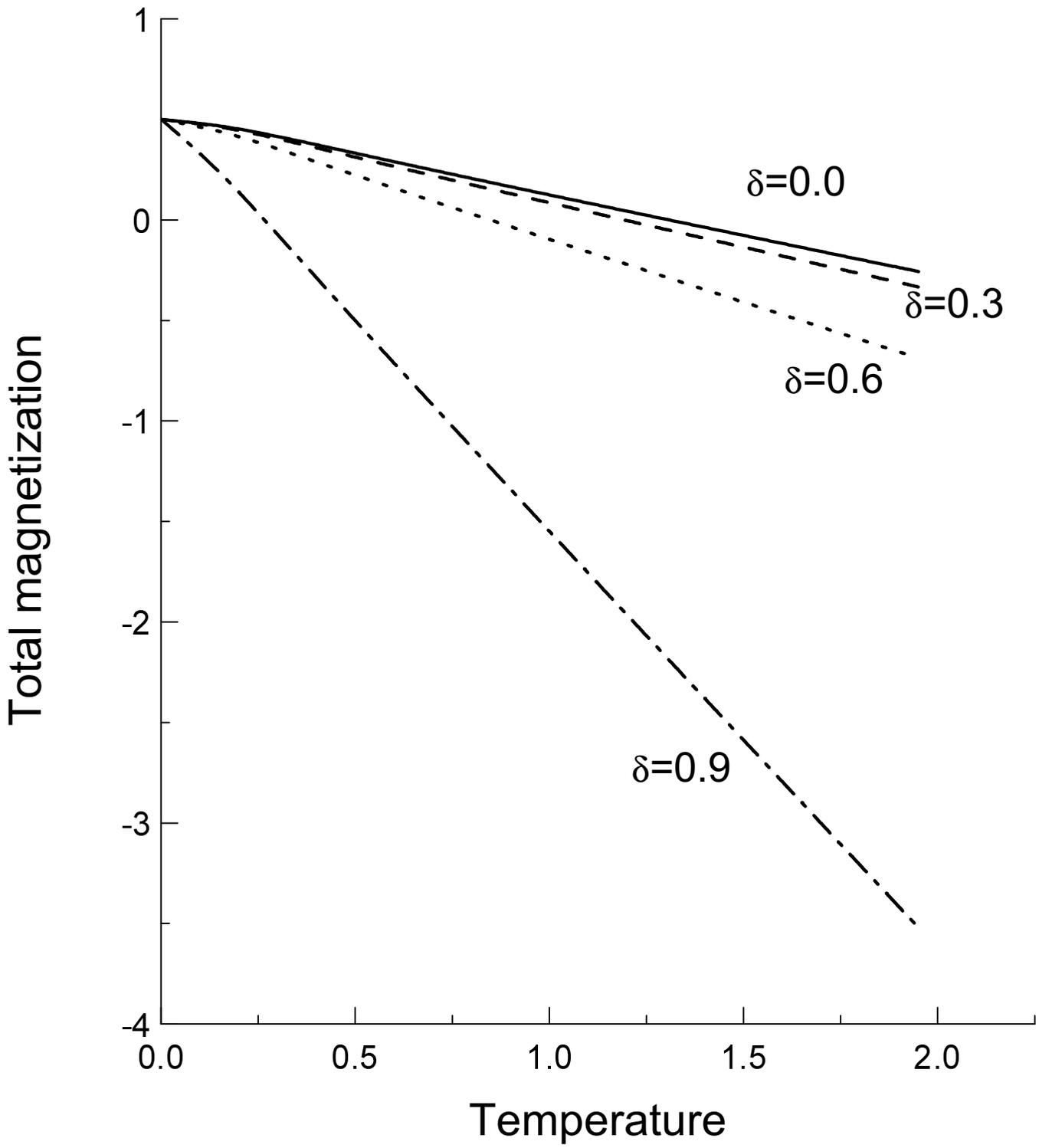
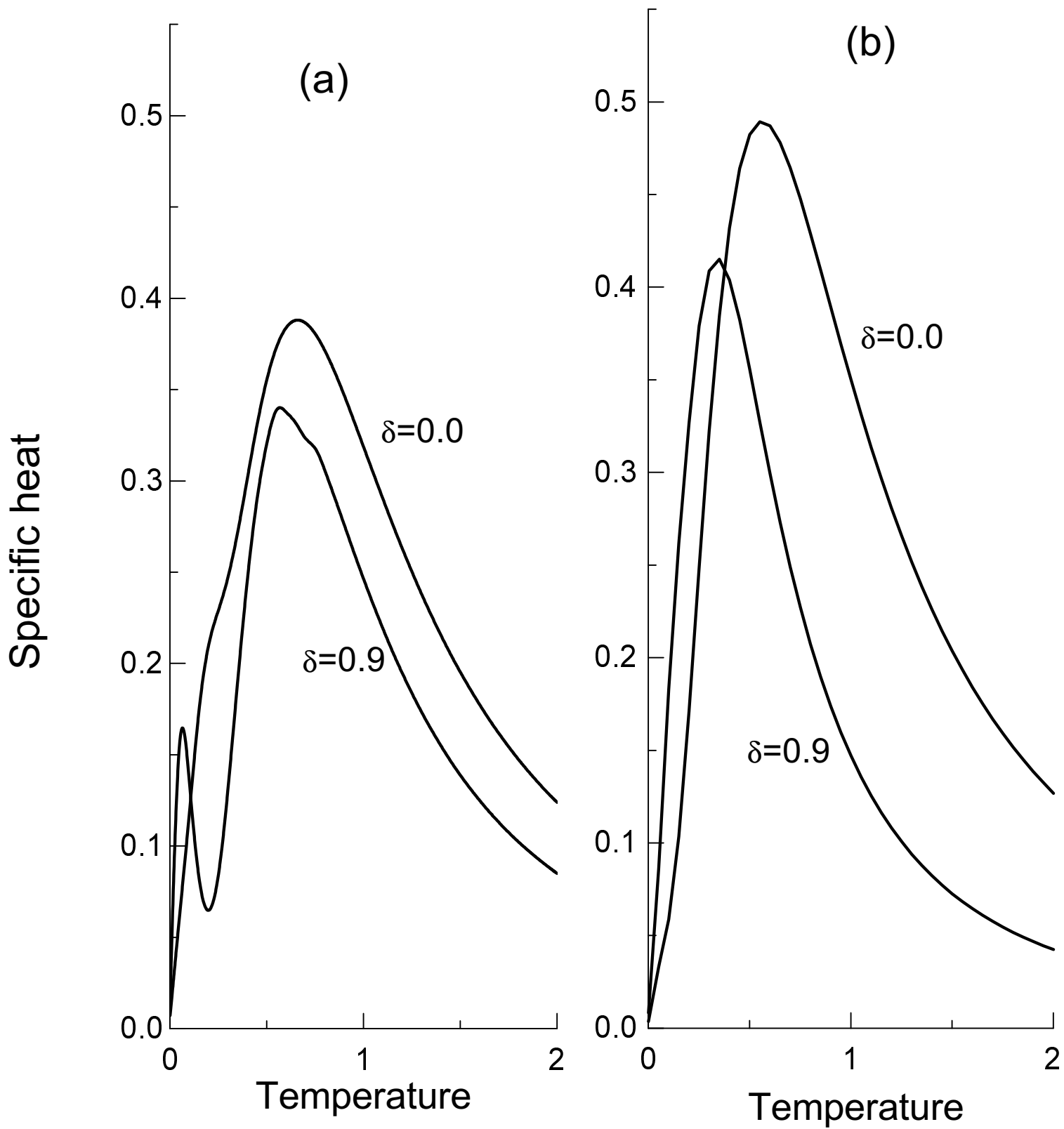
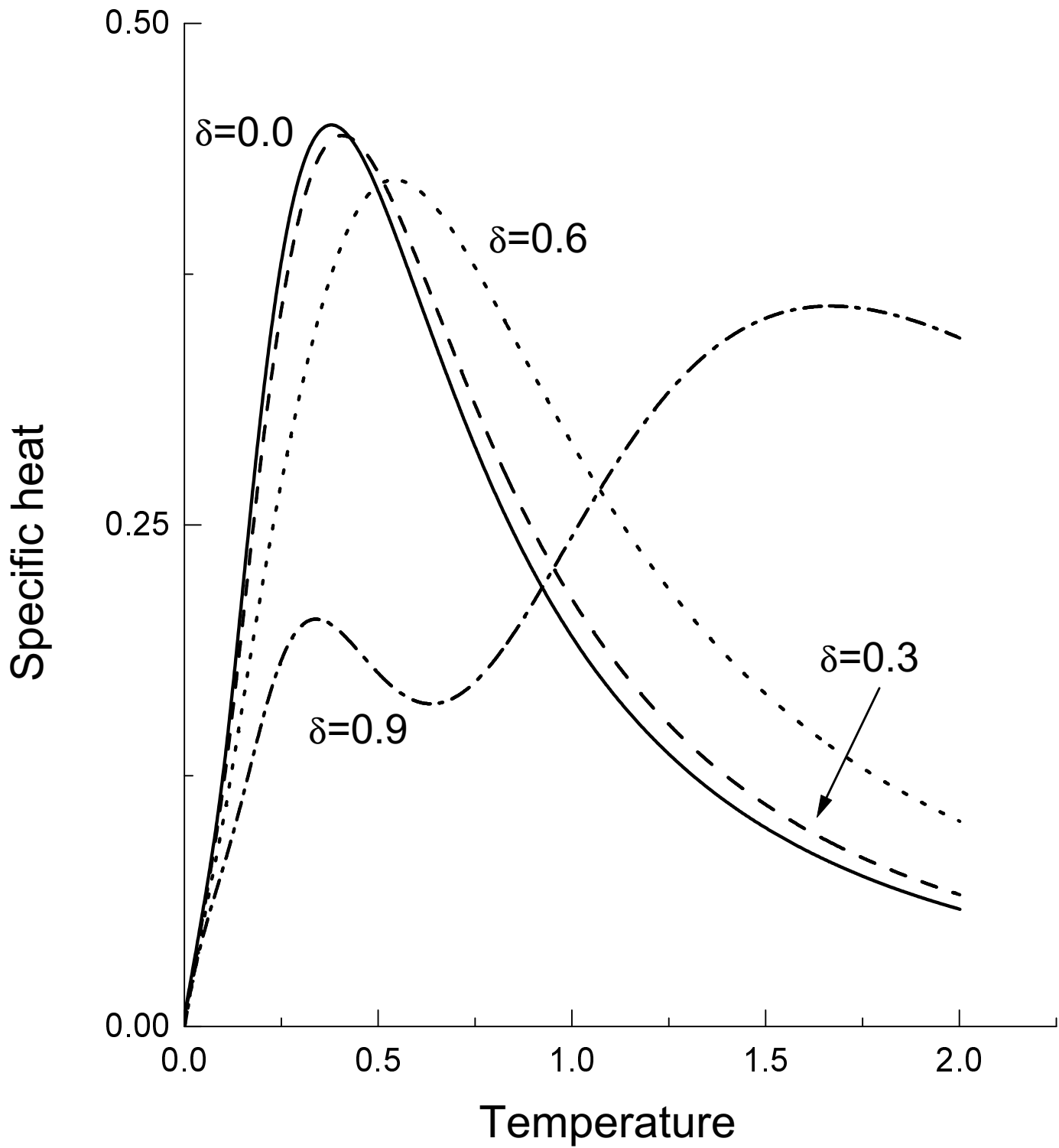
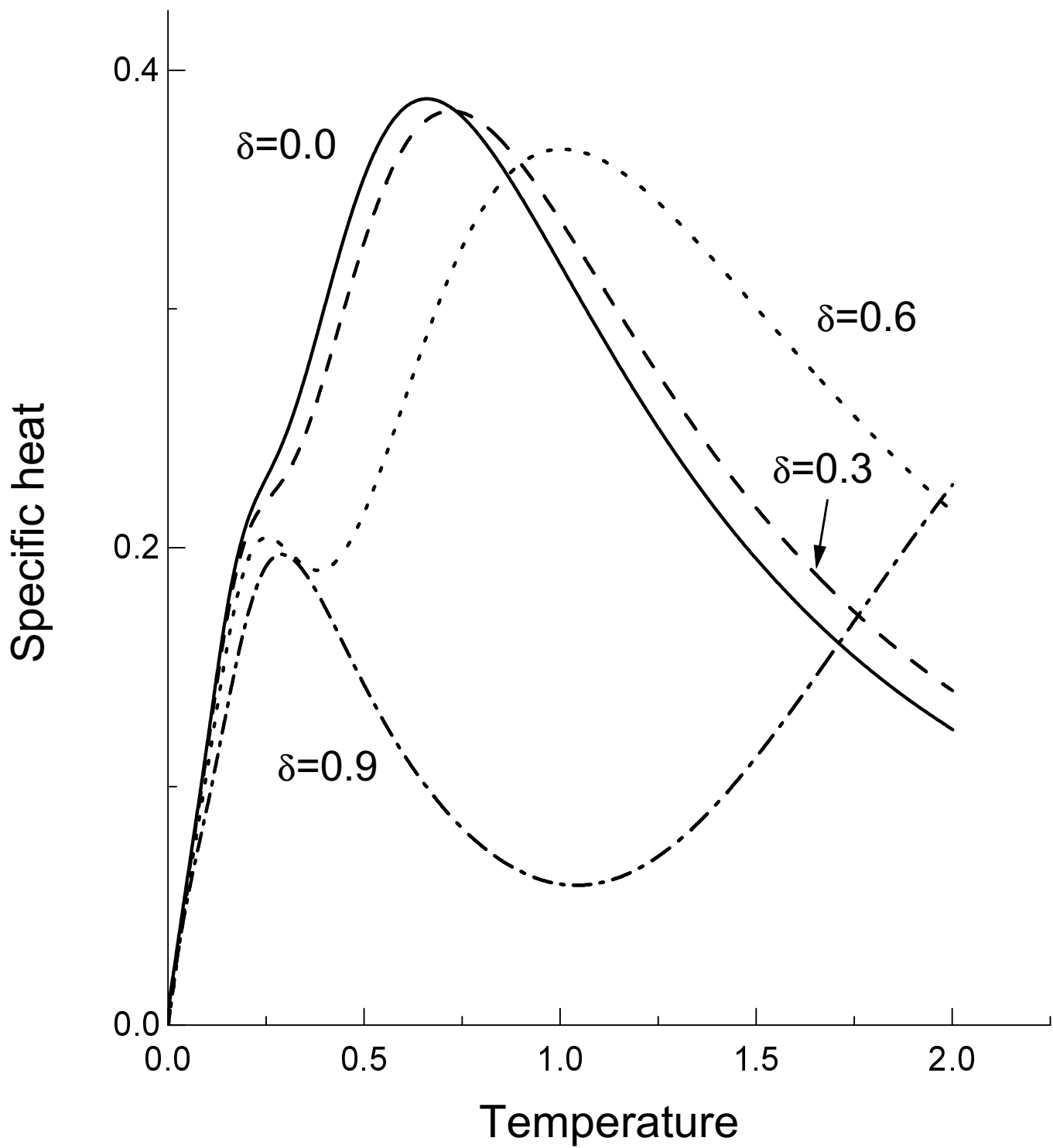
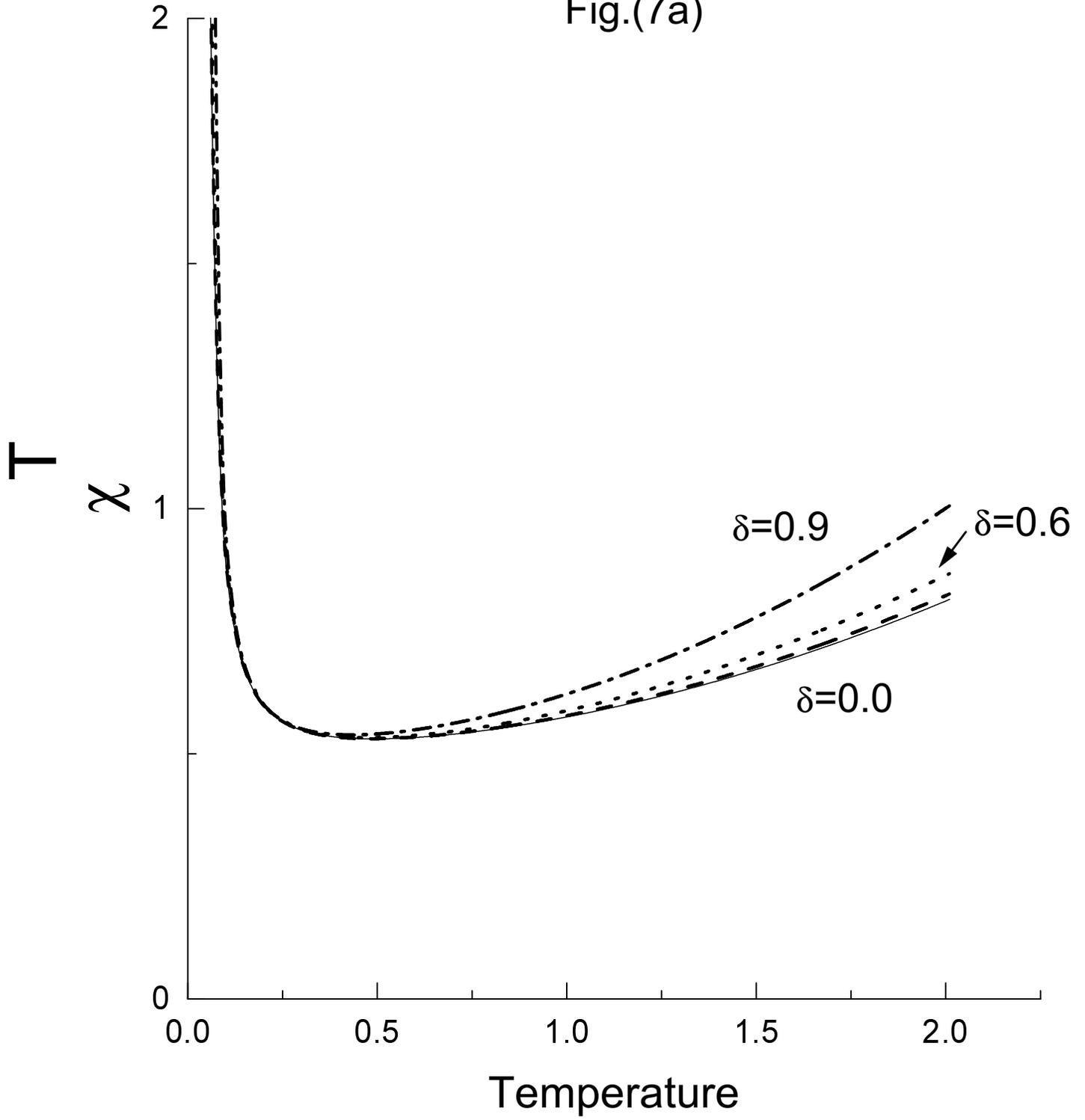
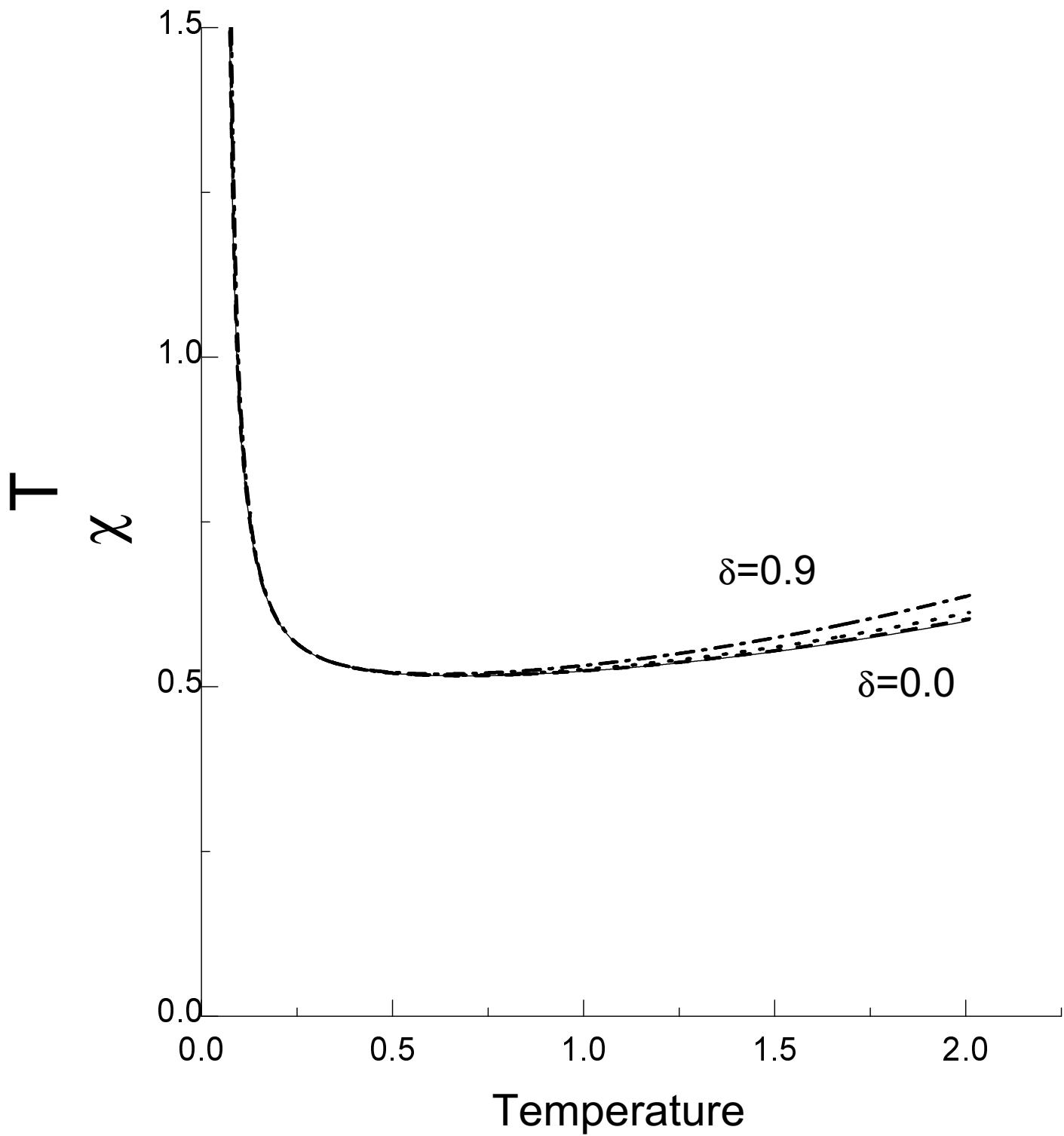
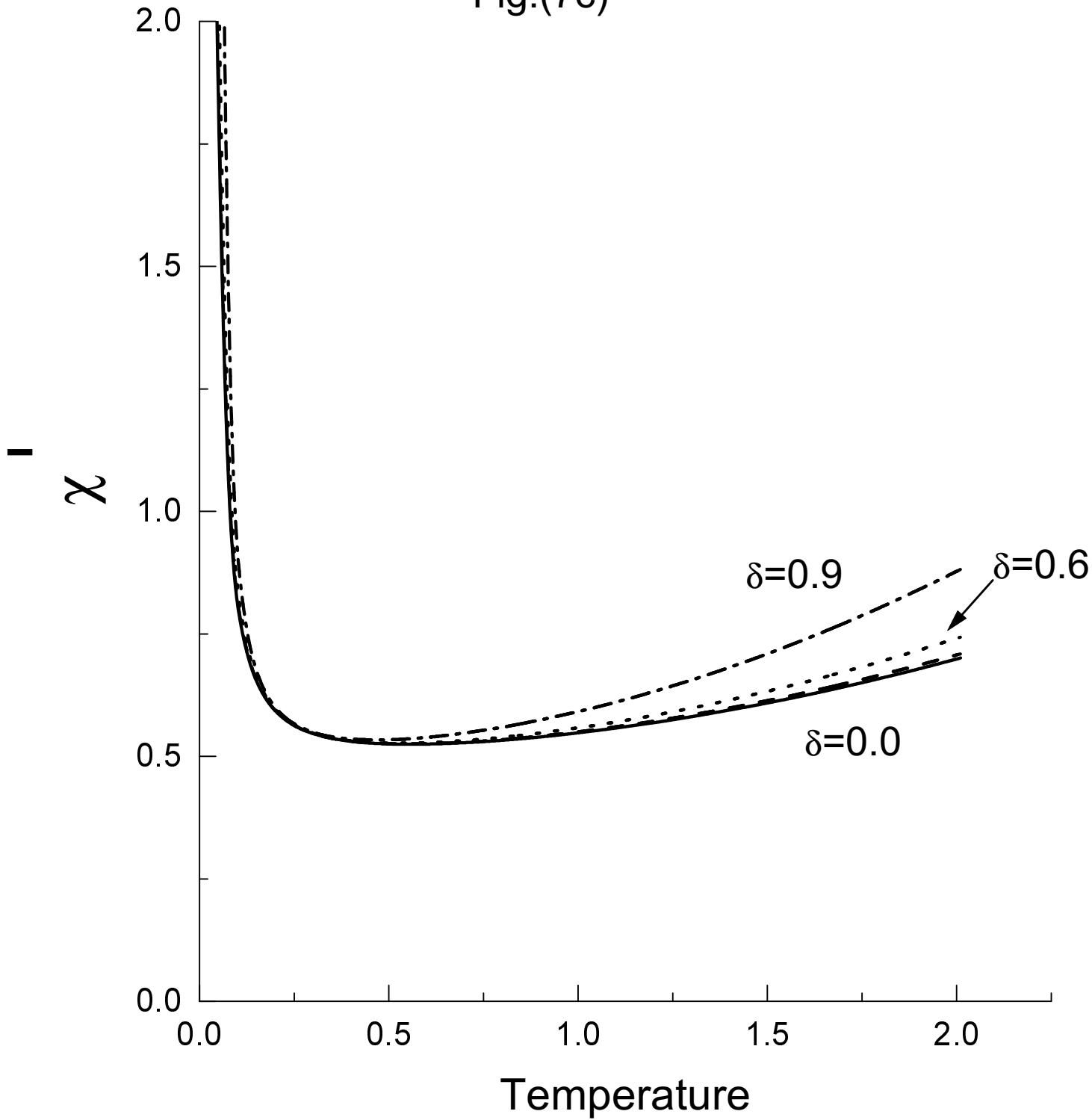
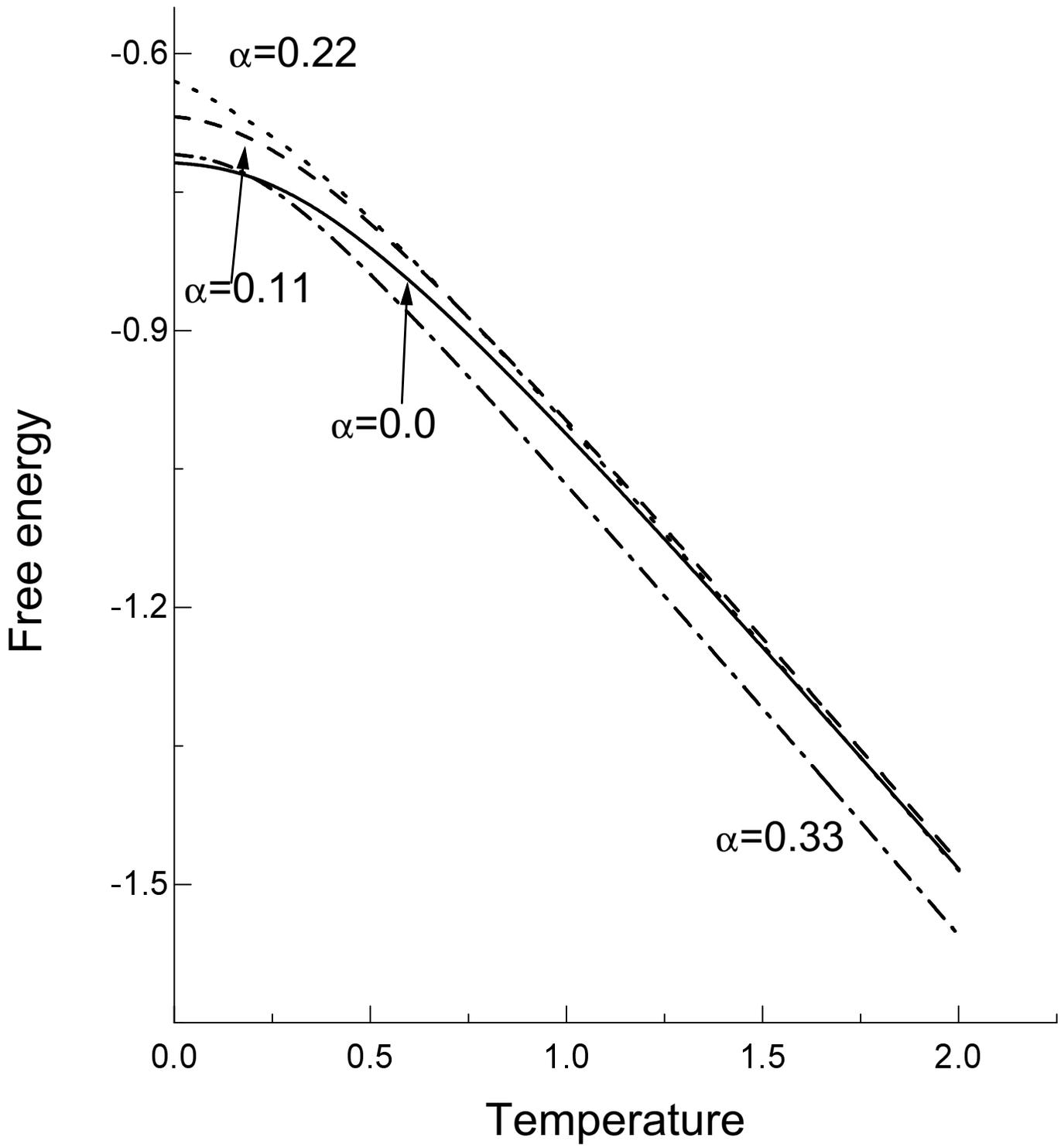
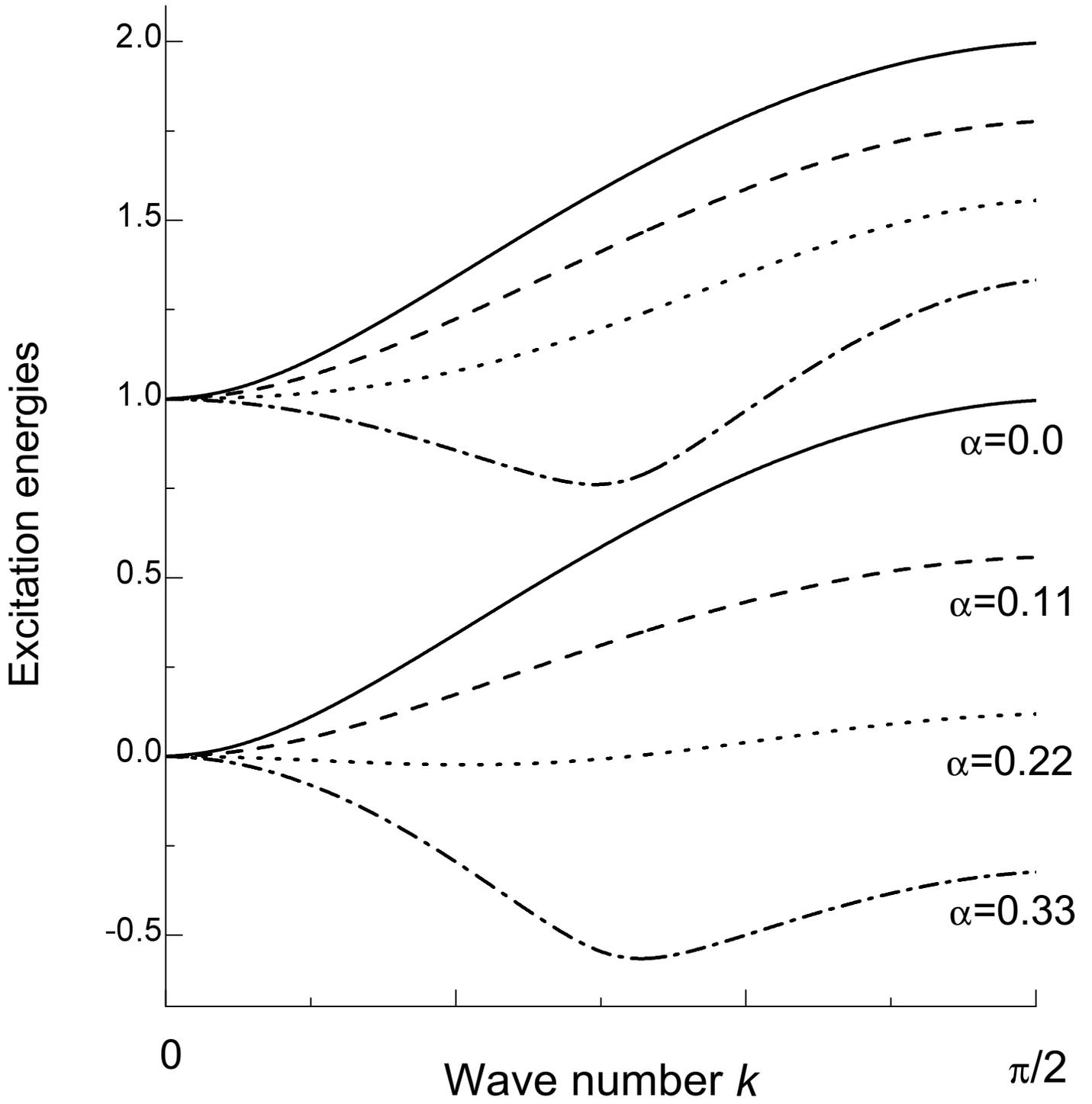
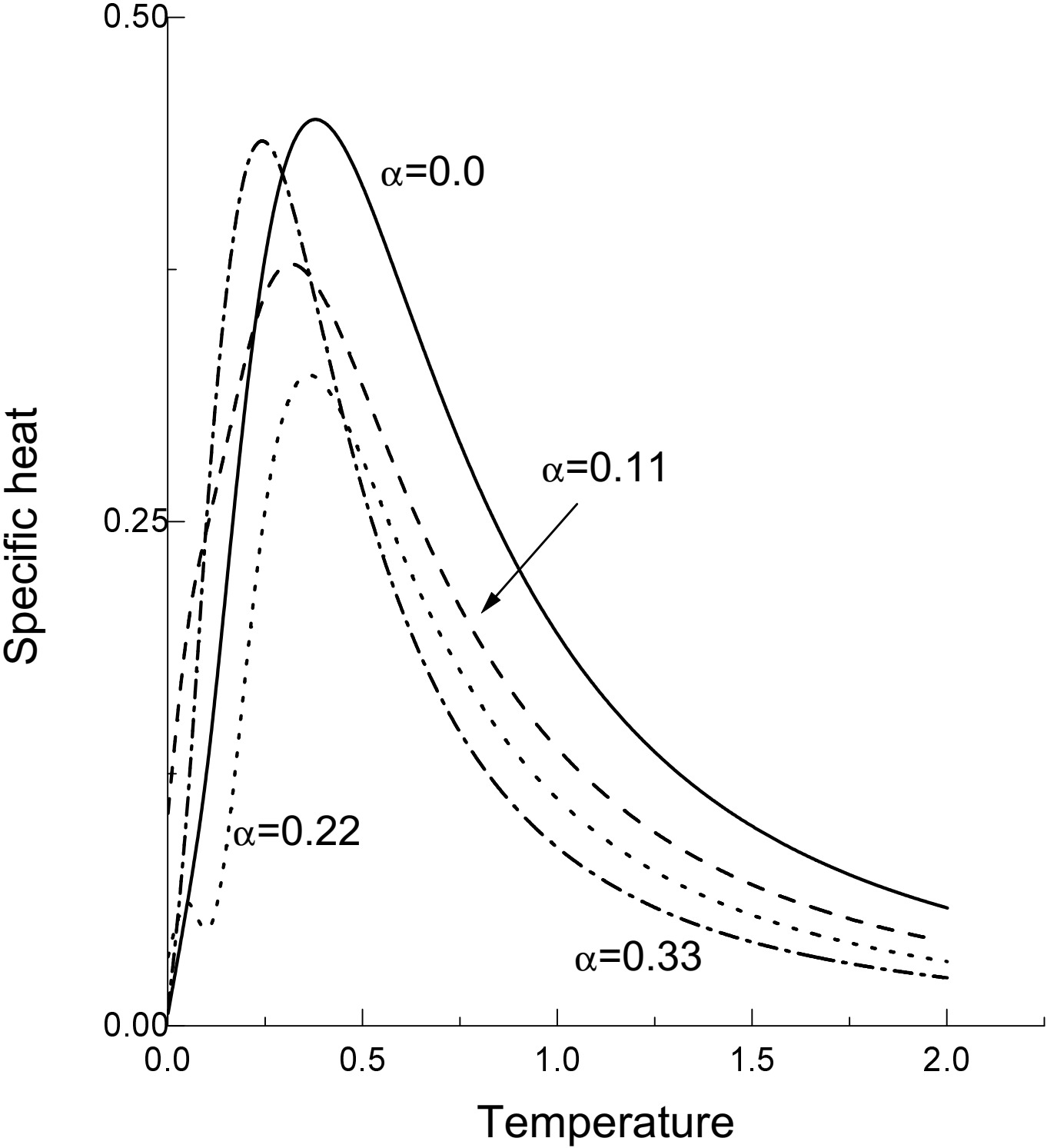
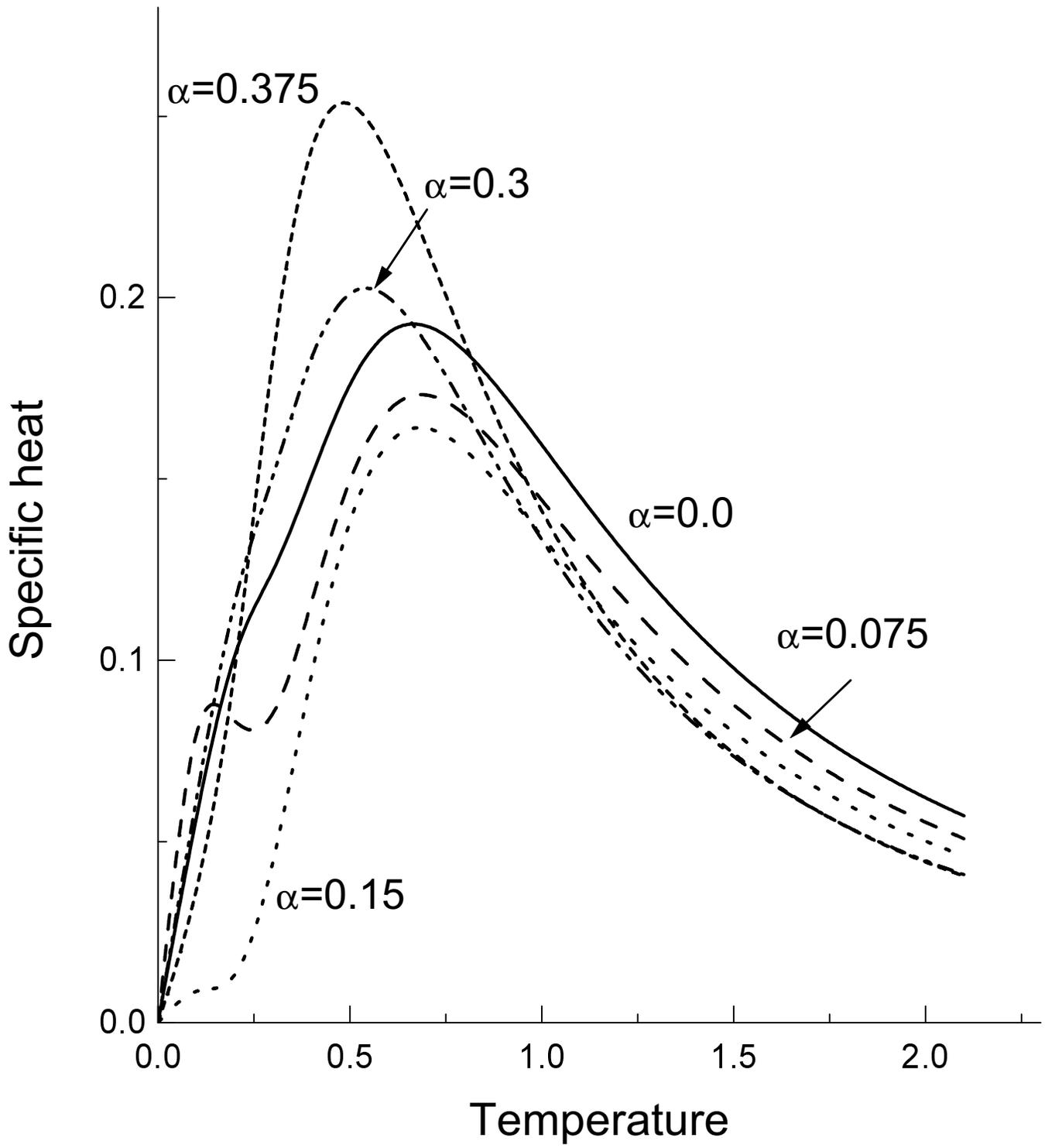
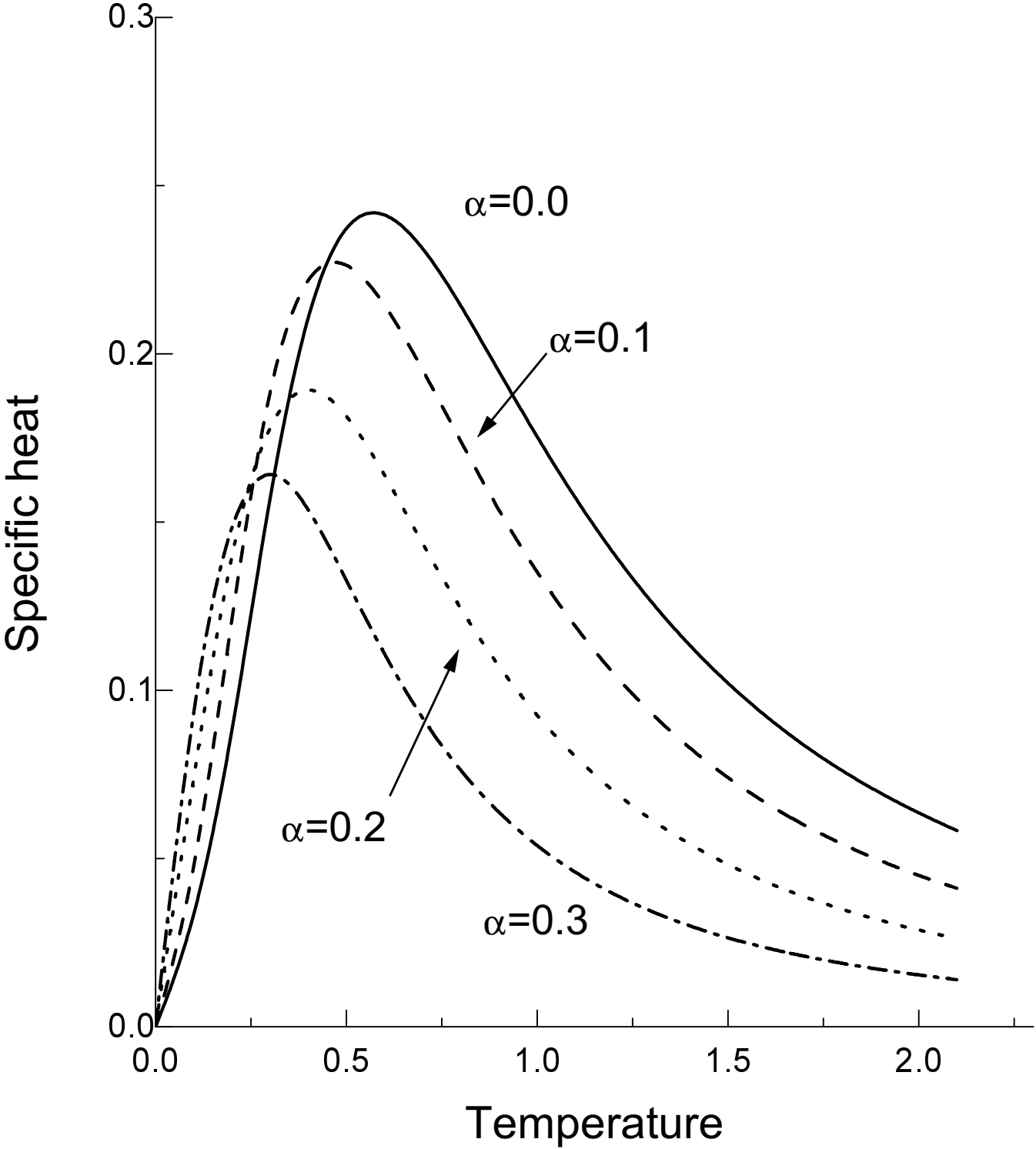
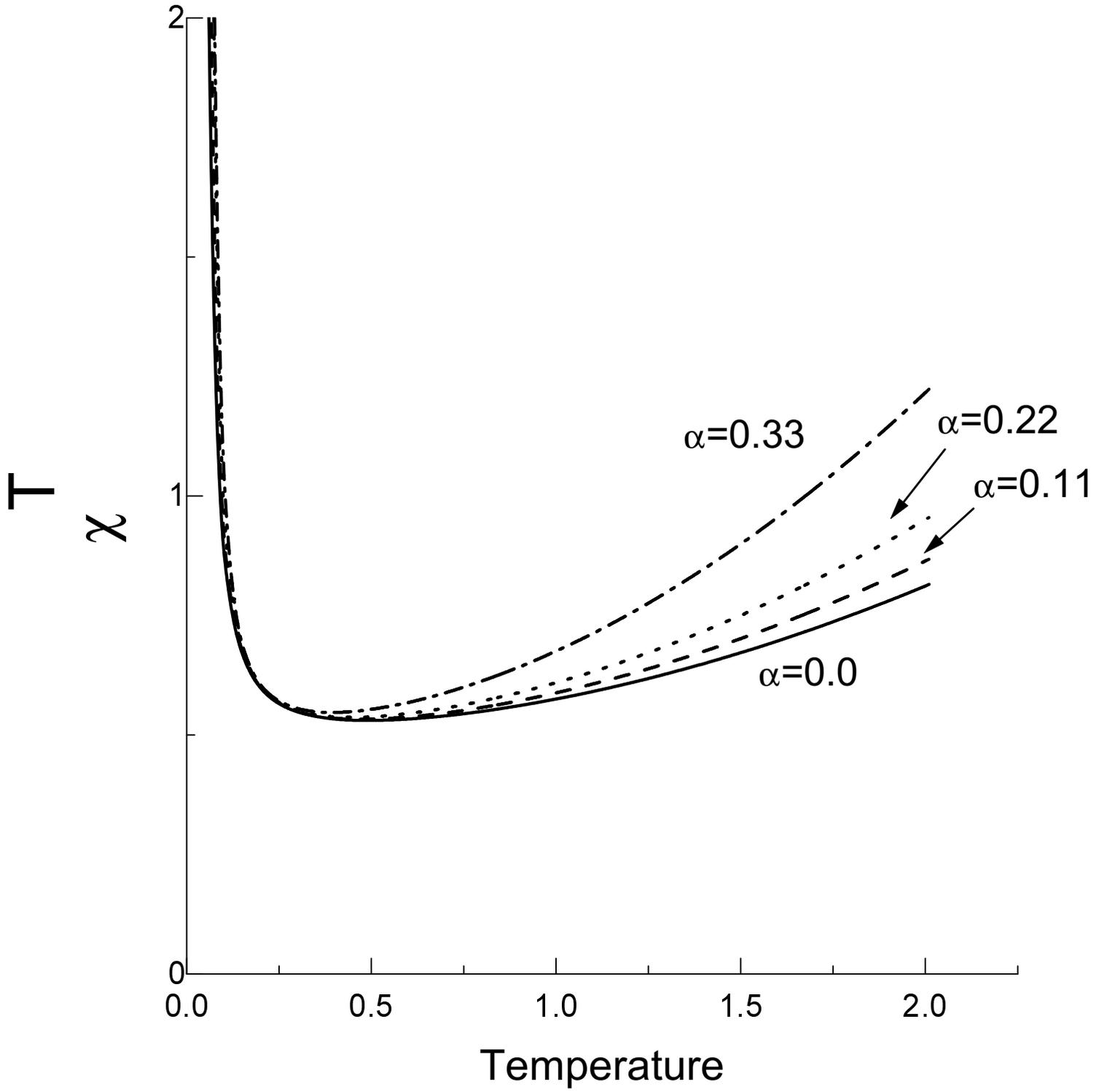
\end{document}